\documentclass[pra,twocolumn,superscriptaddress,showpacs]{revtex4}
\usepackage{amsbsy}
\usepackage{amsfonts}
\usepackage{amssymb}
\usepackage{amsmath}
\usepackage{color}
\usepackage{graphicx}
\usepackage{float}
\usepackage[caption = false]{subfig}
\usepackage{ulem}
\begin{document}


\title{Quantum Critical Probing and Simulation of Colored Quantum Noise}

\author{Eduardo Mascarenhas}
\affiliation{Institute of Physics, Ecole Polytechnique F\'{e}d\'{e}rale de Lausanne (EPFL), CH-1015 Lausanne, Switzerland}

\author{In\'es de Vega}
\affiliation{Physics Department and Arnold Sommerfeld Center for Theoretical Physics,
Ludwig-Maximilians-Universitat Munchen, D-80333 Munchen, Germany}

\begin{abstract}
We propose a protocol to simulate the evolution of a non-Markovian open quantum system by considering a collisional process with a many-body system, which plays the role of an environment. As a result of our protocol the environment spatial correlations are mapped into the time correlations of a noise that drives the dynamics of the open system. Considering the weak coupling limit the open system can also  be considered as a probe of the environment properties. In this regard, when preparing the environment in its ground state, a measurement of the dynamics of the open system allows to determine the length of the environment spatial correlations and therefore its critical properties. To illustrate our proposal we simulate the full system dynamics with matrix-product-states and compare this with the reduced dynamics obtained with an approximated variational master equation. 
\end{abstract}


\maketitle

\section{Introduction}

Quantum simulation was envisioned as a promising innovation to expand our computational capacity beyond classical resources~\cite{Feynman}, but several years were to pass before this inception led to the development of the quantum simulation field~\cite{Steven,Cirac}. The main stream idea is to use a discrete space-time quantum circuit of two-body gates to mimic, as close as possible, the behavior of complex quantum systems that ultimately cannot be efficiently simulated on a classical computer. Experimental developments have brought the notion of a quantum simulator to firmer grounds in different architectures, including trapped-ions~\cite{Porrinha,Monroe,Blatt,Bollinger}, ultra-cold atoms~\cite{Bloch,Bloch2,Kuhr,Gross}, and superconducting circuits~\cite{Koch}. The simulation of the nonequilibrium dynamics of quantum systems coupled to complex environments is receiving increasing attention~\cite{Mataloni,Piilo,Blatt2,Rosenbach,Prior,Vega,Vega2,Sweke,Nov1,Nov2,Cosco}. 
Several proposals have emerged that include an environment producing a classical noise~\cite{Nov1,Nov2} or even a quantum noise~\cite{Mataloni,Piilo,Blatt2,Rosenbach,Prior,Vega,Vega2,Sweke,Nokkala,Cosco} which may therefore yield to dissipation in the open system~\cite{Burgarth}. Based on a space-time discretization, collisional models are a natural route for the simulation of such complex dynamics and to account for non-Markovian effects~\cite{Diosi,Strunz,Giovannetti,McCloskey,Nadja,Nadja2,Nadja3,Osellame,buzek,Mesco,Mesco2}.  

In this theoretical work we propose to simulate the generic dynamics of an open system via a collisional process. We show that a sequence of collisions of the open system with a many body system containing spatial correlations produces the same reduced dynamics as the one of an open system coupled to a structured environment as described via the usual spin-boson model. The simplicity of the underlying quantum circuit relies on the fact that the effect of the many body environment on the system dynamics is encoded into its spatial correlations. Also, these are mapped into effective noise correlation functions playing the same role as the ones  in the standard open system theory~\cite{breuerbook,vega2017}.
Thus, our formalism simulates the dynamics of an open quantum system driven by a colored quantum noise containing non-Markovian correlations. The protocol requires: (i) preparing a many-body system in a state that encodes the desired spatial correlations and (ii) performing two body gates between the system of interest and the many-body environment. We restrict our analysis to ground state preparations of a 1D environment, which allow us to efficiently compute the full system-environment dynamics with matrix product states (MPS)~\cite{Schollwoeck,Verstraete}. 

\begin{figure}
{\includegraphics[width = 2.5in]{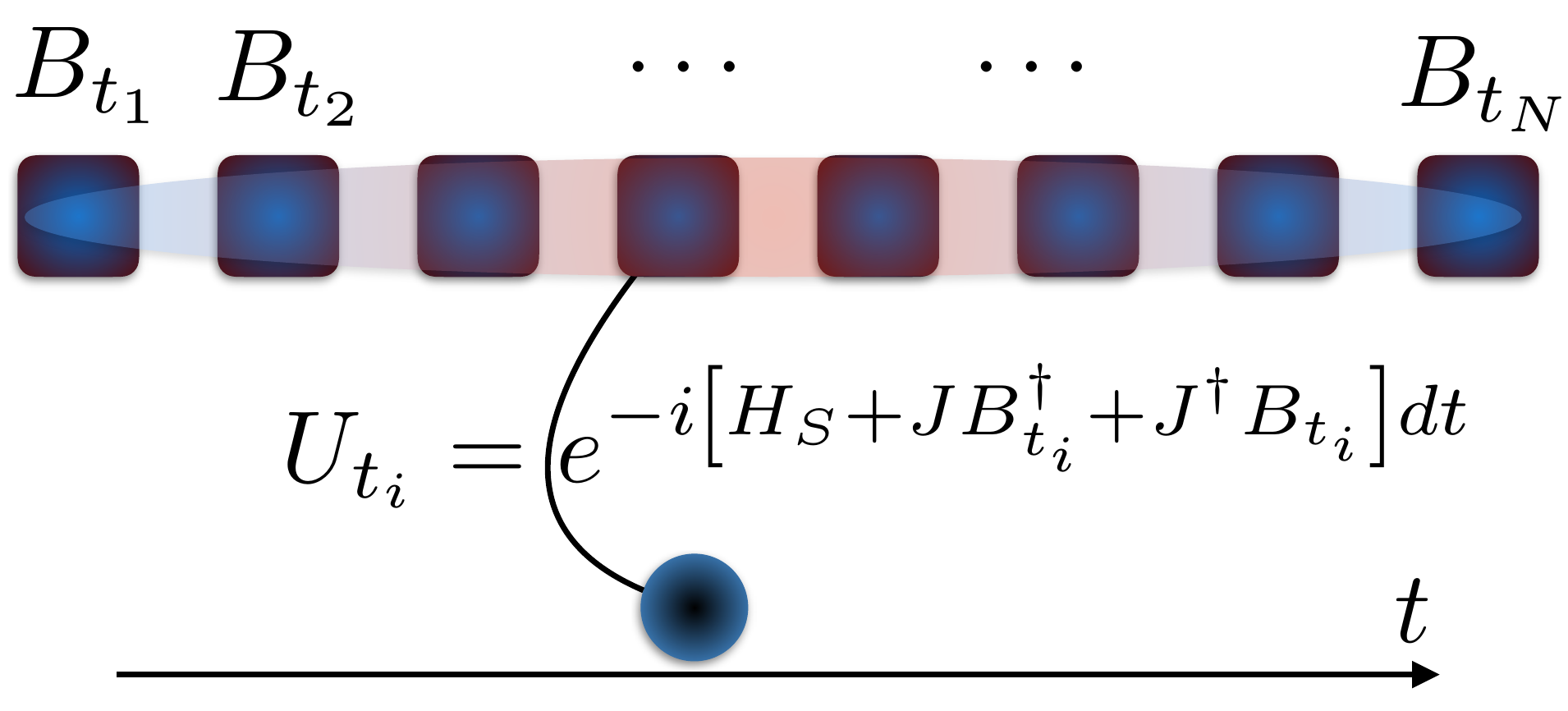}}\\ 
\caption{ Representation of the collisional model from time dependent two-body quantum gates $U_{t_i}$ between the system and the 1D many-body system. At each time step the system moves forward interacting with the next site.}
\label{xfig}
\end{figure}

We note that we are not necessarily interested in simulating a specific non-markovian equation of motion. 
Instead, our focus is to simulate a bath with a specific correlation function.
The resulting weak coupling master equation has a restricted applicability to simple systems. If the system is itself a complex many-body system, we would in principle have to fully diagonalize its hamiltonian in order to derive the closed form master equation. And this is unfeasible for a many-body system in the sense that the resources needed scale exponentially with system size. Therefore, we find and suggest that quantum simulation would be a powerful strategy to describe the dynamics of complex quantum systems interacting with non-white noise baths and also offer a future venue to explore the strong coupling limit in which most master equation formulations are particularly inaccurate.

In the specific case of weak coupling addressed here, we show that the proposed protocol allows to probe criticality of a many body system \cite{Cosco2017,WEVs,WEVs2}. This is done by considering that the open system (i.e. the probe) dynamics is governed by a weak coupling ME that is driven by the environment correlations. Thus, by monitoring the dynamics of a single observable of the probe we can extract properties such as the environment correlation length, which is a typical figure of merit for distinguishing quantum many-body phases. This feature is not present in the previous littelature~\cite{Cosco2017,WEVs,WEVs2}.
We also estimate the back-action that the probe induces on the environment and show that such back-action can be reduced for a sufficiently weak coupling yielding accurate estimation of the correlation length.

\section{The model}

Let us consider a discrete 1D bosonic chain with Hamiltonian $H_B$, described by annihilation (creation) operators $b_i$ ($b_i^\dagger$) located at each site $i$, and prepare it in its ground state. Generically, we define the first order correlations $C^{(1)}(i,j)=\langle b_ib_j^{\dagger}\rangle$, 
$C^{(2)}(i,j)=\langle b_i^{\dagger}b_j\rangle$, $C^{(3)}(i,j)=\langle b_i^{\dagger}b_j^{\dagger}\rangle$, $C^{(4)}(i,j)=\langle b_ib_j\rangle$, 
%
and their corresponding average length,
\begin{equation}\overline{\xi}^{(n)}=\sqrt{\frac{\sum_jj^2C^{(n)}(0,j)}{\sum_jC^{(n)}(0,j)}}.\label{xi2}\end{equation}
In the thermodynamic limit, the ground state phase diagram of a generic many-body system may have non-critical phases with exponentially vanishing correlations $C(i,j)\propto e^{-|i-j|/l}$ and critical phases with power-law correlations $C(i,j)\propto |i-j|^{-K}$~\cite{1DBHDMRG,Carr,Subir,Coleman}.


We now consider a collisional protocol in which a quantum system with Hamiltonian $H_S$ quickly sweeps through the 1D system and weakly interacts with each of its sites one at a time (see illustration in Fig.~\ref{xfig}). We assume that the sweep and quantum gates are fast enough such that the dynamics of the 1D system induced by $H_B$ can be neglected. 
This process is described by a sequence of unitary transformations $U_1(\Delta t),U_2(\Delta t),\cdots,U_i(\Delta t), \cdots,U_N(\Delta t)$ corresponding to two-body gates 
\begin{equation} 
U_i(\Delta t)=e^{-i\left[H_S+H_{\mathrm{int}}(t_i)\right] \Delta t},\label{U}
\end{equation}
acting at collision-times $t_1,t_2,\cdots,t_i,\cdots,t_N$ and lasting for a short time interval $\Delta t=t_{i+1}-t_{i}$. Thus, the collisions occur at times $t_i=i\Delta t $, at positions $x_i=i\Delta x $, and at speed $v=\Delta x/\Delta t$. 
At each collisional time $t_i$ the system-environment coupling is described by an interaction Hamiltonian $H_{\mathrm{int}}(t_i)=Jb_i^{\dagger}\varepsilon+J^{\dagger}b_i\varepsilon$, where $J$ can be identified as a system jump operator, and $\varepsilon$ is the coupling strength. Let us now define the quantum noise  $B_{t_i}=b_i\varepsilon $. This noise processes inherits the correlations of the many-body ground-state such that the static spatial correlations of the 1D environment are perceived by the system as time-correlations of the noise process. Explicitly, we have 
\begin{eqnarray}\langle B_{t_i} B_{t_j}^{\dagger}\rangle &=&\langle b_ib_j^{\dagger}\rangle\varepsilon^2 =C^{(1)}(t_i,t_j) ,\\\nonumber
\langle B_{t_i}^{\dagger} B_{t_j}\rangle&=&C^{(2)}(t_i,t_j),\\\nonumber
\langle B_{t_i}^{\dagger} B_{t_j}^{\dagger}\rangle&=&C^{(3)}(t_i,t_j),\\\nonumber
\langle  B_{t_i} B_{t_j}\rangle&=&C^{(4)}(t_i,t_j),\label{discretecorr}\end{eqnarray}
where we have included the coupling strength in the definition of the correlation functions. 

\begin{figure}
{\includegraphics[width = 3.5in]{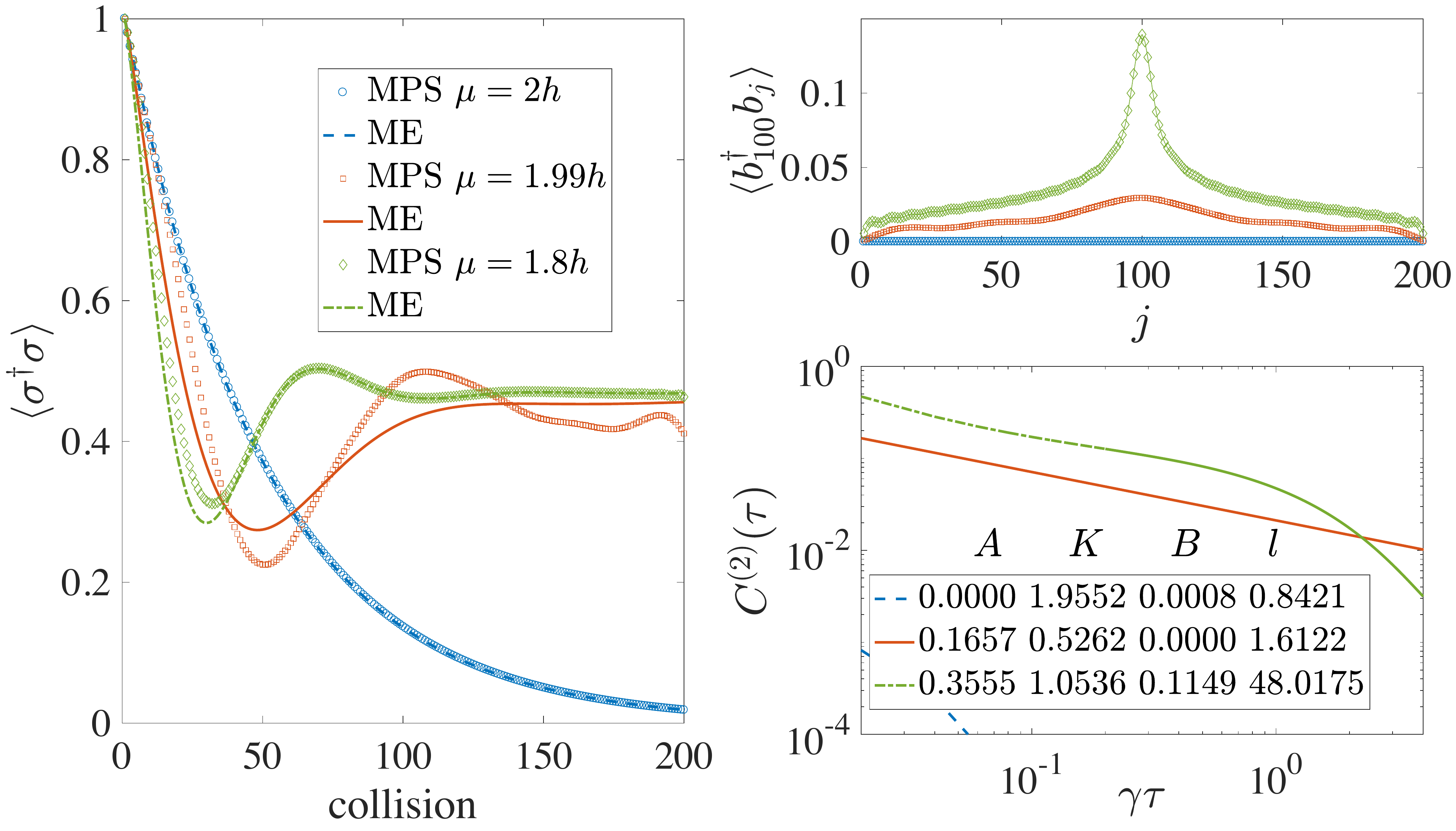}}\\ 
\caption{Data for a BH environment of 200 sites with the q-bit initialized in the up state. (Left) MPS calculated qubit population as a function of collisions with the corresponding ME fit and the (upper right) corresponding ground-state correlations. (Lower right) We also show the fitted $C^{(2)}(\tau)$ in log scale. The blue curve being orders of magnitude smaller thus largely neglected in the plot.}
\label{pop}
\end{figure}

Let us now consider the difference between the system wave function at a time $t+\Delta t$ and the one at a time $t$,
\begin{eqnarray}|\Psi(t+\Delta t)\rangle-|\Psi(t)\rangle= \left[U_t(\Delta t)-1\right]|\Psi(t)\rangle,
\end{eqnarray}
and divide such equation by $\Delta t$. In the following, we assume the continuous limit in which $\Delta t\rightarrow 0$, such that the discrete time and space coordinates become continuous variables, i.e. $t_i\rightarrow t$ and $b_i\rightarrow b_t$, with $t=x/v$ being the continuous limit of the space coordinate and $v$ is the speed of the moving quantum system. 
Thus, in the continuous limit, the evolution resulting from the repeated application of the gates defined in~(\ref{U}) corresponds to the following quantum stochastic Schr\"odinger equation~\cite{QNoise} 
\begin{eqnarray}\frac{d|\Psi (t)\rangle}{dt}&=&-i\left[H_S+H_{\mathrm{int}}(t)\right]|\Psi(t)\rangle, 
\label{QSSE}
\end{eqnarray}
where $H_{\mathrm{int}}(t)=JB^\dagger_t+J^{\dagger}B_t$, where $B_t=b_t \varepsilon$. In interaction picture with respect to the system, we have $d|\tilde{\Psi} (t)\rangle=-i\tilde{H}_{\mathrm{int}}(t)|\tilde{\Psi}(t)\rangle dt$, with $\tilde{H}_{\mathrm{int}}(t)=J_tb_t^{\dagger}\varepsilon+J_t^{\dagger}b_t\varepsilon$, and $J_t=e^{iH_S t}Je^{-iH_S t}$. 
Considering the interaction picture, a ME for the reduced state of the system $\tilde{\rho}^{(S)}=\mathrm{tr_{B}}\big\{ \tilde{P}_t
\big\}$, with $\tilde{P}_t=|\tilde{\Psi} (t)\rangle\langle \tilde{\Psi} (t)|$ may be derived as
\begin{equation}\frac{d\tilde{\rho}_t^{(S)}}{dt}=-i\mathrm{tr_{B}}\left\{\big[J_t B_t^{\dagger}+J_t^{\dagger}B_t,\tilde{P}_t\big]\right\},\end{equation}
which requires computing averages such as $\mathrm{tr_{B}}\big\{ B_t\tilde{P}_t\big\}$, $\mathrm{tr_{B}}\big\{ B^\dagger\tilde{P}_t\big\}$, $\mathrm{tr_{B}}\big\{ \tilde{P}_t B_t\big\}$ and $\mathrm{tr_{B}}\big\{ \tilde{P}_t B^\dagger\big\}$. In order to compute them up to second order in the coupling parameter $\varepsilon$, we perform a perturbative expansion of the projector $\tilde{P}_t$, 
\begin{eqnarray}
\tilde{P}_t=\tilde{P}_0-i\int_0^t dt' [H_{\mathrm{int}}(t'),\tilde{P}_0]+{\mathcal O}(\varepsilon^2).
\label{pt_pert}
\end{eqnarray}
where we consider that $\tilde{P}_0=P_0=\rho_0^{(S)}\otimes \rho_B(0)$, with $\rho_0^{(S)}$ and $\rho_B(0)$ the system and environment initial states, respectively.
Thus, we find a Novikov-like relation~\cite{Nov1,Nov2,Nov3} valid up to second order
\begin{multline}\mathrm{tr_{B}}\big\{ B_t\tilde{P}_t\big\}=-i\langle B_t\rangle \rho_0^{(S)}
-i\int_0^t dt'\Big[\langle B_t B_{t'}\rangle J^\dagger_{t'}\tilde{\rho}_t^{(S)}\\
+\langle B_t B_{t'}^{\dagger}\rangle J_{t'}\tilde{\rho}_t^{(S)}-\tilde{\rho}_t^{(S)}J^\dagger_{t'}\langle B_t B_{t'}\rangle
-\tilde{\rho}_t^{(S)}J_{t'}\langle B_t B^\dagger_{t'},
\rangle \Big], \label{Novikov}\end{multline}
and a similar equation is found for $\mathrm{tr_{B}}\big\{ B^\dagger\tilde{P}_t\big\}$, $\mathrm{tr_{B}}\big\{ \tilde{P}_t B_t\big\}$ and $\mathrm{tr_{B}}\big\{ \tilde{P}_t B^\dagger\big\}$. Here, we have defined $\langle B_t\rangle=\mathrm{tr_{B}}\{B_t \rho_B(0)\}$ and $\langle B_t B_{t'}^{\dagger}\rangle=\mathrm{tr_{B}}\{ B_t B_{t'}^{\dagger}\rho_B(0)\}$. Also, consistently to second order we have replaced $\rho_0^{(S)}\approx\tilde{\rho}_t^{(S)}$, with $\tilde{\rho}_t^{(S)}=e^{iH_S t}\rho_t^{(S)}e^{-H_S t}$,  in all the second order terms at the right hand side of the equation (\ref{Novikov}). In addition, the time correlation functions as the continuum limit of their discrete counterparts given by Eqs. (\ref{discretecorr}). 
Considering this, and going back to the Schr\"odinger equation, we find that the master equation up to second order in the coupling parameter can be written as 
\begin{eqnarray}
&&\frac{d\rho_t^{(S)}}{dt}=-i[H_S,\rho^{(S)}_t]-i[J\langle B^\dagger_t\rangle+J^\dagger\langle B_t\rangle,\rho^{(S)}_0]\cr
&-&\frac{1}{2}\int_0^tdt'\Big\{C^{(1)}(t,t')\left[ J^{\dagger}J_{t'-t}\rho^{(S)}_{t'}-J_{t'-t}\rho_{t'}^{(S)}J^{\dagger}+\mathrm{h.c.} \right] \cr
&+&C^{(2)}(t,t')\left[ J_{t'-t}J^{\dagger}\rho^{(S)}_{t'}-J^{\dagger}\rho_{t'}^{(S)}J_{t'-t}+\mathrm{h.c.} \right]\cr
&+&C^{(3)}(t,t')\left[ J_{t'-t}J\rho^{(S)}_{t'}-J_{t'-t}\rho_{t'}^{(S)}J+\mathrm{h.c.} \right]\cr
&+&C^{(4)}(t,t')\left[ J_{t'-t}^{\dagger}J^{\dagger}\rho^{(S)}_{t'-t}-J_{t'-t}^{\dagger}\rho_{t'}^{(S)}J^{\dagger} +\mathrm{h.c.}\right]\Big\}.
\label{ME}\end{eqnarray}
As it can be seen, this ME is identical to the one of a standing open system coupled with a strength $g_k$ to a set of independent harmonic oscillators (characterized by $b_k$ ($b_k^\dagger$) and having eigenfrequencies $\omega_k$ and a state $\rho_B$), as described with the spin-boson model. This model leads to correlations of the form $C^{(1)}(t,t')=\sum_{k,k'} g_kg_{k'} \mathrm{tr_{B}}\{\rho_B b_{k'}b_{k}^\dagger\}e^{i\omega_k t-i\omega_{k'}t'}$, for instance.
 
Moreover, to probe the environment state we shall measure the reduced dynamics of the open system to get $\rho^M_{t'}=\mathrm{tr_B}\left\{ |\Psi_{t'}\rangle\langle\Psi_{t'}| \right\}$ governed by the eq.~(\ref{QSSE}). Further, considering that this quantity is also approximately obtained with the ME~(\ref{ME}) we may employ a variational optimization to determine the parameters in the correlations $C^{(l)}(t,t')$ that best minimize the distance
\begin{equation}\min_{\{ C^{(n)}\}}\int_0^t\left|\mathrm{tr_B}\left\{ |\Psi_{t'}\rangle\langle\Psi_{t'}| \right\} -\rho^{(S)}_{t'}\right|dt'.\label{Var}\end{equation}

\begin{figure}
{\includegraphics[width = 3.in]{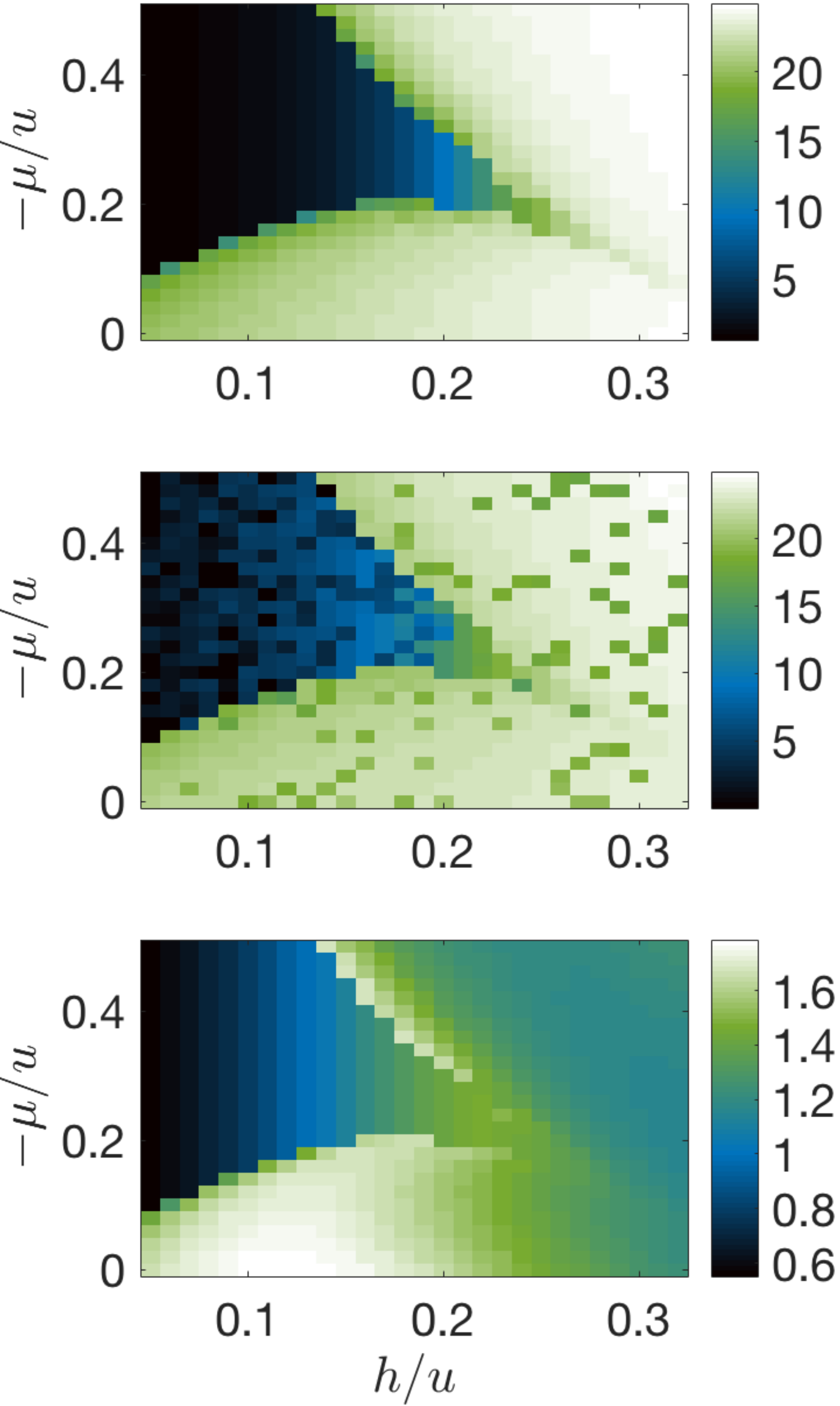}}\\ 
\caption{ (Top) Direct MPS calculated correlation length $\xi^{(2)}$, (middle) the same quantity variationally obtained from the corresponding ME, and (bottom) the probe population $\langle a^{\dagger}a\rangle$ after the last collision for a Bose-Hubbard system of 50 sites with collision time steps of $dt=0.01/\gamma$. The probe is always initialized in the empty or vacuum state.}
\label{BHGround50}
\end{figure}

\section{Dissipation of a qubit}

 As a first benchmark we show that the proposed protocol allows to simulate decaying dynamics, contrary to proposals based on classical noise~\cite{Burgarth}. 
We consider that the system is a qubit having negligible free dynamics for simplicity. We also consider as environment a 1D Bose-Hubbard  
\begin{equation}H_{\mathrm{B}}=\sum_i \left[-h(b_ib^{\dagger}_{i+1}+b_i^{\dagger}b_{i+1})+\frac{u}{2}b_i^{\dagger}b_i^{\dagger}b_ib_i +\mu b_i^{\dagger}b_i\right]. \end{equation}
Here, $h$ is the hopping rate, $u$ is the on site interaction between bosons and $\mu$ is a local energy scale or chemical potential.
In the thermodynamic limit, the ground state phase diagram of the model has a Mott phase with exponentially vanishing correlations and a critical superfluid phase with power-law correlations~\cite{1DBHDMRG,Carr}. 
The system is coupled to the environment via a jump operator $J=\sqrt{\gamma}\sigma$, that is, a lowering operator, with $\gamma$ being the effective coupling strength to the environment. Using SWAP gates and an MPS ansatz we simulate the protocol of the traveling qubit at constant speed as it traverses the 1D environment with $N=200$ oscillators, by integrating eq.~(\ref{QSSE}). The maximal truncation error we have is $10^{-11}$, the bond dimension is $D=500$ and the environment local dimension is $d=5$. 
We also analyze the system dynamics with the approximate ME~(\ref{ME}). Given our choice for the environment, at very low densities we may assume that the correlations governing the equation have the form
 \begin{equation}C^{(2)}(i,j)=A(1+|i-j|)^{-K}+Be^{-|i-j|/l},\end{equation}
with $C^{(1)}=C^{(2)}$ and $C^{(1)}(0)=1+C^{(2)}(0)$ and $C^{(3)}=C^{(4)}=0$ as corresponds to Gaussian colored noise. Thus, the resulting ME~(\ref{ME}) describes the dissipative decay of an open system coupled to an environment in equilibrium. Further, the effects of temperature could be included by adding collisions with a second lattice, following a thermofield transformation \cite{Vegathermo}. 

In Fig.~\ref{pop} we show the time evolution of the qubit population and the correlations of the environment ground-state for both MPS and ME results. We observe pure exponential decay for a non-critical environment at $\mu=2h$. Changing the chemical potential leads to a transition into the longer-range correlated (superfluid) states. The resulting system dynamics present more structure, as it corresponds to a non-Markovian regime. This is the opposite case of what is observed in Ref. \cite{Cosco} for a standing probe, where the superfluid regime leads to more Markovian dynamics. We also point out that the MPS based system-environment simulation is exact, as it takes into account fine-size effects and the back-action that the system exerts on the environment, while the approximate ME neglects both. Thus, mismatches are expected even though we find very good agreement between the two models. Very close to the phase transition the mentioned back-action seems more dramatic leading to the worst match. Slightly farther away from the phase transition the ME seems to be a very accurate approximation.

\section{A bosonic quantum probe}

We now explore higher density regimes, more specifically a region around the first Mott lobe. In addition, we use a bosonic system having a jump operator $J=\sqrt{\gamma}a$ with annihilation operator $a$. The probing aspect of our scheme is shown in Fig.~\ref{BHGround50} in which we report an overview of the phase diagram around the first Mott lobe. In detail, we show both the correlation length (\ref{xi2}) directly calculated from the ground-state and extracted from the probe. The latter result is obtained by using a generic search algorithm~\cite{GA} to solve a simplified version of the variational problem of Eq. ~(\ref{Var}),  
\begin{equation}\min_{\{ C^{(n)}\}}\int_0^t\left|\mathrm{tr}\left\{O |\Psi_{t'}\rangle\langle\Psi_{t'}| \right\} -\mathrm{tr}\left\{O\rho^{(S)}_{t'}\right\}\right|dt',\label{Var2}\end{equation}
with $O=a^{\dagger}a$. The advantage of this simplification is that while Eq.~(\ref{Var}) would require the tomography of the probe state in an experimental setting, Eq.~(\ref{Var2}) relies solely on the dynamics of a single observable. Comparing the top and middle panels of Fig.~\ref{BHGround50} we can see that the probed correlation length is faithful to the original. There are, however, small fluctuations in the optimization procedure. This is due to the fact the the variational problem has very ``shallow" minima such that it is numerically difficult to resolve within a small vicinity around the optimal solution. Interestingly, the bottom panel of Fig.~\ref{BHGround50} shows that the probe population after the last collision (approaching its steady state) undergoes a transition which resembles the environment phase transition. We shall remark that even though the correspondence between the top and bottom panel is remarkable it is not perfect since the environment size is relatively small, 50 sites, and therefore the probe dynamics approaches but does not reach the steady state of the map that emerges from the collision process. 

\begin{figure}
{\includegraphics[width = 3.4in]{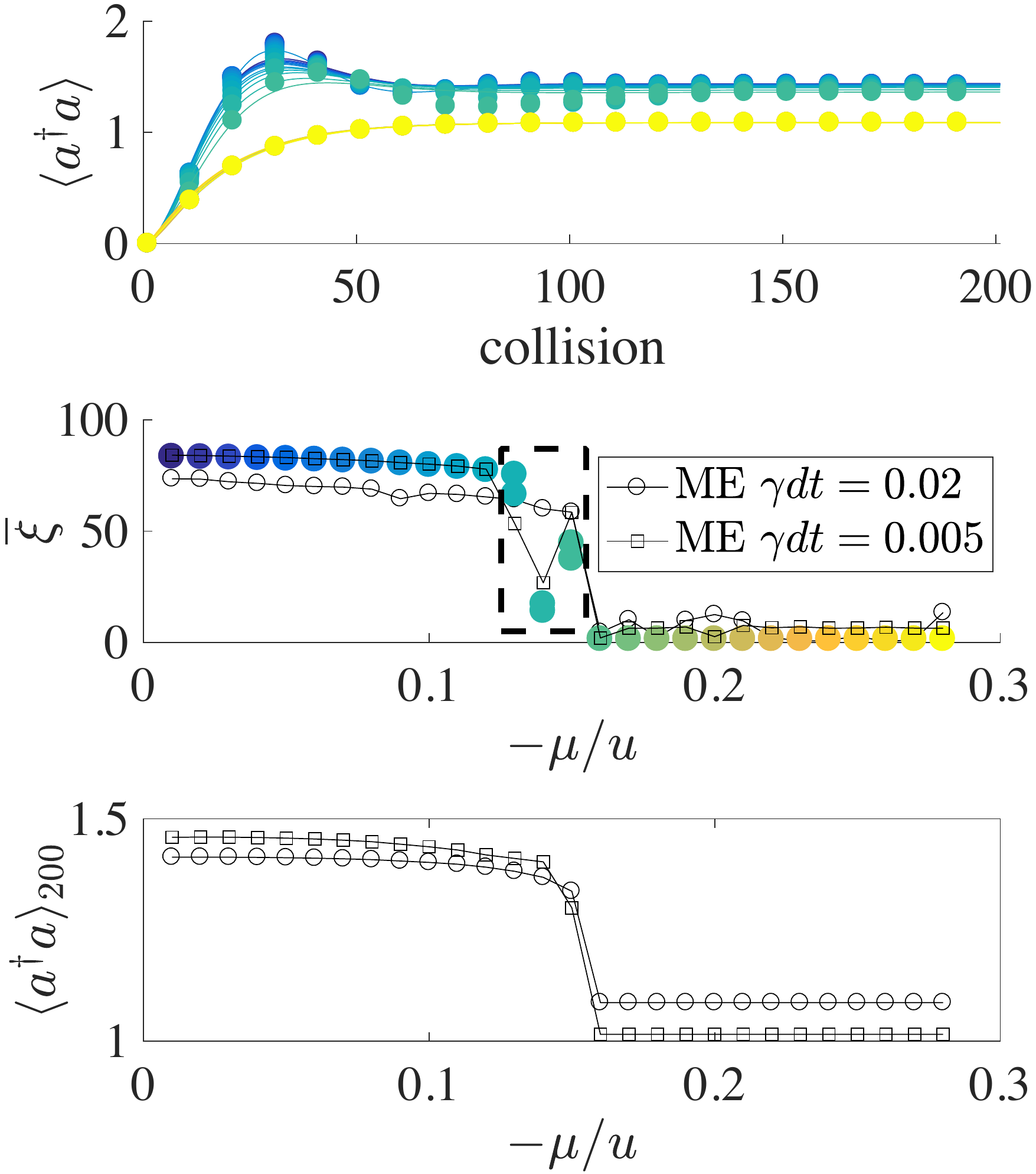}}\\ 
\caption{Bosonic probing of a Bose-Hubbard system of 200 sites for $h=0.1u$ and probe initialized in vacuum. (Top panel) Population of the probe calculated with MPS (markers) and ME (solid lines) as a function of collisions and considering $\gamma dt=0.02$. Darker and lighter colors correspond to smaller and larger values of $-\mu/u$ respectively, which are specified in the middle panel of the same figure. 
(Middle panel) Correlation lengths both for the unperturbed ground-state (full-markers) and probed result as given by the ME (empty markers). (Bottom) Asymptotic population of the probe after 200 collisions, with the symbols having the same interpretation as in the middle panel.}
\label{popLobe}
\end{figure}

Next we analyse the dynamics of the probe, its steady state and the system back-action into the environment state. To this aim, we increase the environment size to 200 sites. In the top panel of Fig.~\ref{popLobe} we show the dynamics of the probe for a collision strength (given by the rate between the interaction time and the system decay time scale $\sim 1/\gamma$), $\gamma dt=0.02$.
Lighter colors correspond to larger $-\mu/u$ and thus fall inside the Mott phase that generates monotonic (exponential) dynamics. Darker colors fall inside the super-fluid phase that generates structured dynamics.  The middle panel represents the correlation length  $\xi^{(2)}$ computed from the MPS calculation of the ground state (full colored markers), and the variational ME (empty markers) considering again two values of collision strength $\gamma dt$. When considering strong collisions, the error in estimating the original correlation length is bigger in the Mott phase than in the superfluid phase. However, for weaker collisions $\gamma dt=0.005$ we find virtually perfect agreement in all regimes. The dashed rectangle in the middle panel indicates the transition region in which it becomes numerically challenging to converge with MPS to the ground state of such large systems. More specifically, all our variational determinations of the ground states have converged with 5 MPS steps in the variational algorithm except at this region. 
In the bottom panel of Fig.~\ref{popLobe} we show that the asymptotic population of the probe after 200 collisions shows a strong signature of the phase transition that confirms that the ground state phase transition appears to produce a dynamical phase transition on the probe. 

Our algorithm could be simulated with ultracold atoms of two types, $a$ and $b$, which can be achieved by considering atoms in two different hyperfine ground states. Similar to the proposals in Refs.  \cite{recati2005,devega2008}, atoms in $a$ and $b$ correspond to the open system and its environment respectively. To this purpose, atoms in $a$ are trapped in a deep optical lattice that keeps them in a Mott state where only the first level of each lattice well is involved. Hence, the atomic dynamics within each well of such lattice are described by the ladder operator $J$ ($J^\dagger$) that represents transitions between a state where there is an atom in the well and a state where there are no atoms. In turn, atoms in $b$ are trapped by a tunable optical lattice that implements the Bose-Hubbard physics in the standard way \cite{Cirac}. 

\begin{figure}
{\includegraphics[width = 2.5in]{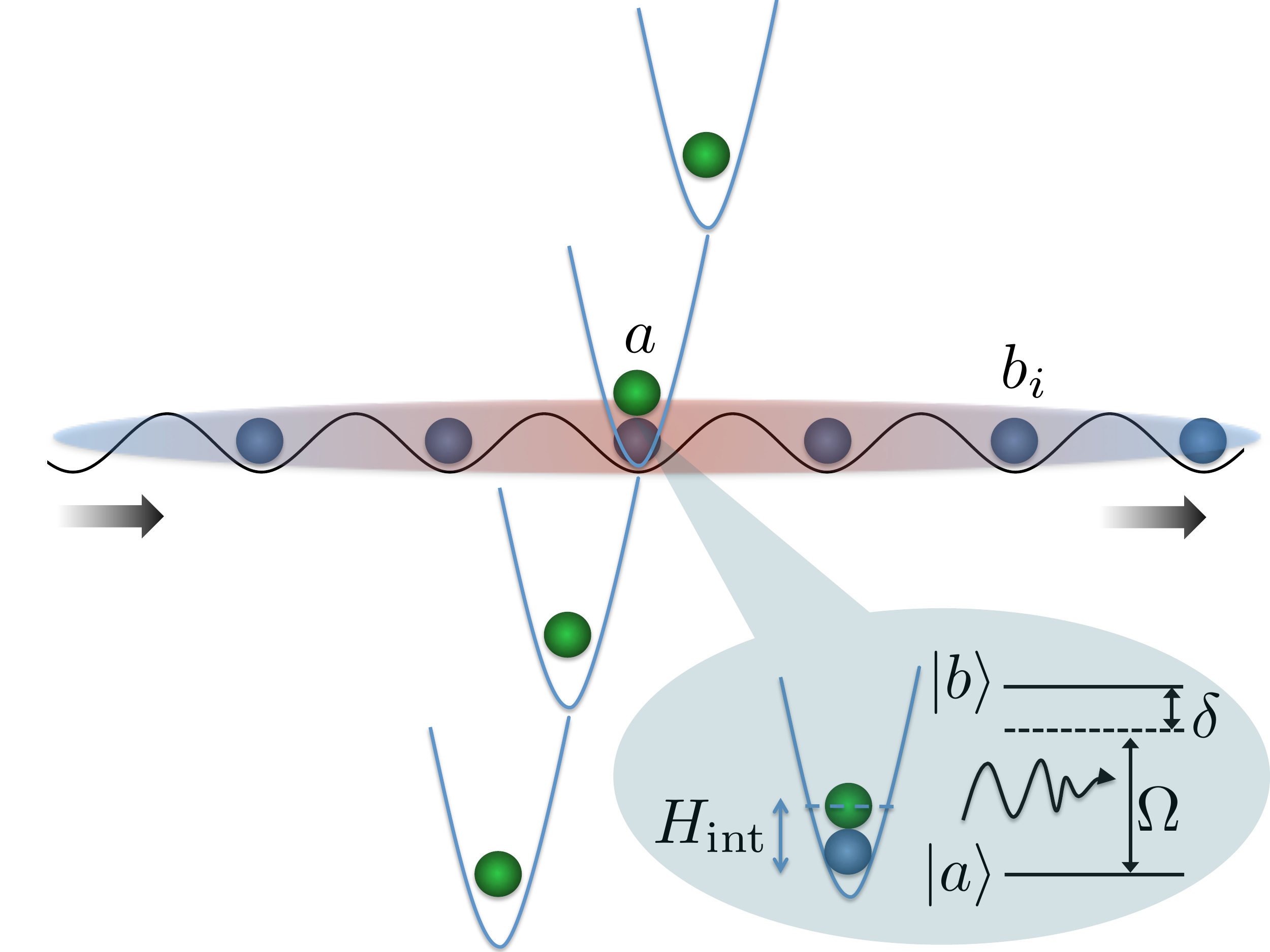}}\\ 
\caption{ Possible experimental implementation of the collisional model with ultra-cold atoms.}
\label{EXP}
\end{figure}

The two optical lattices are located perpendicular to each other with a single crossing point,  where the interaction between the two types of atoms occur, as sketched in Fig.~\ref{EXP}. 
In order to perform the protocol, the lattice that traps $b$ atoms, for instance, is sequentially shifted with respect to the $a$ lattice by dynamically tuning the corresponding lasers. Such lattice shifting was theoretically proposed in \cite{Jaksch} and first experimentally realized in \cite{bloch2003}. In that way, a single site $a$ sequentially interacts with each site $b_i$ at a time, as described by the interaction Hamiltonian $Jb_i^{\dagger}\varepsilon\Delta t +J^{\dagger}b_i\varepsilon\Delta t$. 
The interaction strength $\varepsilon$ as well as the interaction time $\Delta t$ can be experimentally tuned. 

The nature of the interaction, and therefore the interaction strength depends on the choice of the implementation. For instance, in \cite{recati2005} the interaction is produced by combining a laser that couples the two hyperfine levels $a$ and $b$ with a collisional process. Such a collisional process is described by a contact pseodupotential with coupling parameter $g_{ab}=4\pi a_{ab}\hbar^2/m$ ($a_{ab}$ the corresponding s-wave scattering length and $m$ is the atomic mass) that determines the coupling strength between the system and the environment. Following the proposal in \cite{devega2008}, a second possibility is to consider that atoms in $b$ are coupled to atoms in $a$ only through a two photon Raman transition. In this case, the coupling strength of the interaction Hamiltonian is simply determined by the laser Rabi frequency $\Omega$, i.e.  $\epsilon\sim \Omega$, and therefore is also completely tunable. As an alternative to the above schemes, one may consider an impurity with two internal levels immersed in a three dimensional BEC, as proposed in \cite{marino2017}.

We shall remark that in the present protocol the other sites in the lattice $a$ do not come into play, since such atoms are assumed to be in a Mott insulating phase where no tunneling to neighboring sites is allowed. Allowing the tunneling in lattice $a$ (or even shifting such a lattice too) would nevertheless lead to an interesting interplay between the many-body dynamics of atoms in $a$ and the dissipation produced by the sequential coupling with atoms in $b$.

\section{Conclusion}

We have proposed a collisional model for simulating colored quantum noise, which is based on a simple quantum circuit and does not rely on multiple collisions inside the bath. The formalism allows us to harness the emerging correlations of many-body ground-states to generate non-Markovian dynamics. Conversely, we have shown that the protocol may be used to probe quantum phases via their correlations also showing that the probe back-action can be made negligible. Ultimately, the protocol can be used as a basis to implement non-Markovian dynamics of many-body open systems also at strong coupling, which may require colliding many-body systems (having arbitrary dimension and structure).

{\it Acknowledgements}: We would like to thank Vincenzo Savona and Ulrich Sch\"olwock for encouragement and support, and A. Recati for useful discussions regarding the implementation. I. de V. further acknowledges support by the DFG-grant GZ: VE 993/1-1.


\begin{thebibliography}{99}

\bibitem{Feynman} Richard Feynman, ``Simulating Physics with Computers", International Journal of Theoretical Physics. 21 (6-7): 467 (1982).

\bibitem{Steven} Tomi H. Johnson, Stephen R. Clark and Dieter Jaksch, ``What is a quantum simulator?", EPJ Quantum Technology 1:10 (2014).

\bibitem{Cirac} J. Ignacio Cirac and Peter Zoller, ``Goals and opportunities in quantum simulation", Nature Physics 8, 264?266 (2012).

\bibitem{Porrinha} A. Friedenauer, H. Schmitz, J. T. Glueckert, D. Porras and T. Schaetz, ``Simulating a quantum magnet with trapped ions", Nature Physics 4, 757 - 761 (2008).

\bibitem{Monroe} K. Kim, M.-S. Chang, S. Korenblit, R. Islam, E. E. Edwards, J. K. Freericks, G. D. Lin, L. M. Duan and C. Monroe, ``Quantum simulation of frustrated Ising spins with trapped ions", Nature 465, 590?593 (2010).

\bibitem{Blatt} B. P. Lanyon, C. Hempel, D. Nigg, M. Muller, R. Gerritsma, F. Zahringer, P. Schindler, J. T. Barreiro, M. Rambach, G. Kirchmair, M. Hennrich, P. Zoller, R. Blatt, C. F. Roos, ``Universal digital quantum simulation with trapped ions", Science 334, 57 (2011).

\bibitem{Bollinger} Joseph W. Britton, Brian C. Sawyer, Adam C. Keith, C. C. Joseph Wang, James K. Freericks, Hermann Uys,	Michael J. Biercuk and John J. Bollinger, ``Engineered two-dimensional Ising interactions in a trapped-ion quantum simulator with hundreds of spins", Nature 484, 489?492 (26 April 2012).

\bibitem{vega2017} I.de vega and Daniel Alonso, "Dynamics of non-Markovian open quantum systems",Rev. Mod. Phys., 89, 015001 (2017).

\bibitem{Bloch} Immanuel Bloch, Jean Dalibard and Sylvain Nascimbene, ``Quantum simulations with ultracold quantum gases", Nature Physics 8, 267?276 (2012).

\bibitem{Bloch2} Markus Greiner, Olaf Mandel, Tilman Esslinger , Theodor W. Hansch and Immanuel Bloch, ``Quantum phase transition from a superfluid to a Mott insulator in a gas of ultracold atoms", Nature 415, 39-44 (2002).

\bibitem{Kuhr} Jacob F. Sherson, Christof Weitenberg, Manuel Endres, Marc Cheneau, Immanuel Bloch and Stefan Kuhr, ``Single-atom-resolved fluorescence imaging of an atomic Mott insulator", Nature 467, 68?72 (2010). 

\bibitem{Gross} Martin Boll, Timon A Hilker, Guillaume Salomon, Ahmed Omran, Jacopo Nespolo, Lode Pollet, Immanuel Bloch, Christian Gross, ``Spin-and density-resolved microscopy of antiferromagnetic correlations in Fermi-Hubbard chains", Science Vol. 353, Issue 6305, pp. 1257 (2016).

\bibitem{Koch} Andrew A. Houck, Hakan E. Tureci and Jens Koch, ``On-chip quantum simulation with superconducting circuits", Nature Physics 8, 292 (2012).

\bibitem{buzek} Tom�� Ryb�r and Sergey N Filippov and M�rio Ziman and Vladim�r Bu�ek, "Simulation of indivisible qubit channels in collision models", J. Phys. B: At. Mol. Opt. Phys. 45,154006 (2012).

\bibitem{Mataloni} Andrea Chiuri, Chiara Greganti, Laura Mazzola, Mauro Paternostro and Paolo Mataloni, ``Linear Optics Simulation of Quantum Non-Markovian Dynamics", Scientific Reports 2, Article number: 968 (2012).

\bibitem{Piilo} Bi-Heng Liu, Li Li, Yun-Feng Huang,	Chuan-Feng Li,	Guang-Can Guo, Elsi-Mari Laine, Heinz-Peter Breuer and Jyrki Piilo, ``Experimental control of the transition from Markovian to non-Markovian dynamics of open quantum systems", Nature Physics 7, 931?934 (2011).

\bibitem{Blatt2} Julio T. Barreiro, Markus Muller,	 Philipp Schindler, Daniel Nigg, Thomas Monz, Michael Chwalla, Markus Hennrich, Christian F. Roos,	Peter Zoller and Rainer Blatt, ``An open-system quantum simulator with trapped ions", Nature 470, 486?491 (2011).

\bibitem{Rosenbach} Robert Rosenbach, Javier Cerrillo, Susana F. Huelga, Jianshu Cao and Martin B. Plenio, ``Efficient simulation of non-Markovian system-environment interaction", New J. Phys. 18 023035 (2016).

\bibitem{Prior} Javier Prior, Alex W. Chin, Susana F. Huelga, and Martin B. Plenio, ``Efficient Simulation of Strong System-Environment Interactions", Phys. Rev. Lett. 105, 050404 (2010).

\bibitem{Vega} Ines de Vega, ``Lattice mapping for many-body open quantum systems and its application to atoms in photonic crystals", Phys. Rev. A 90, 043806 (2014).


\bibitem{Vega2} Ines de Vega, Ulrich Schollwock, and F. Alexander Wolf, ``How to discretize a quantum environment for real-time evolution", Phys. Rev. B 92, 155126 (2015).

\bibitem{Sweke} Ryan Sweke, Ilya Sinayskiy, Denis Bernard, and Francesco Petruccione, ``Universal simulation of Markovian open quantum systems", Phys. Rev. A 91, 062308 (2015).

\bibitem{Nov1} A. Chenu, M. Beau, J. Cao, and A. del Campo, ``Quantum Simulation of Generic Many-Body Open System Dynamics Using Classical Noise", Phys. Rev. Lett. 118, 140403 (2017).

\bibitem{Nov2} J. I. Costa-Filho, R. B. B. Lima, R. R. Paiva, P. M. Soares, W. A. M. Morgado, R. Lo Franco, and D. O. Soares-Pinto, ``Enabling quantum non-Markovian dynamics by injection of classical colored noise", Phys. Rev. A 95, 052126 (2017).

\bibitem{Nokkala} J. Nokkala, F. Galve, R. Zambrini, S. Maniscalco, J. Piilo, ``Complex quantum networks as structured environments: engineering and probing", Sci Rep. 6, 26861 (2016).

\bibitem{Cosco} F. Cosco and M. Borrelli, J.J. Mendoza-Arenas, F. Plastina, D. Jaksch and S.  Maniscalco, "Bose-Hubbard lattice as a controllable environment for open quantum systems", arXiv preprint 1706.09148.

\bibitem{Cosco2017} F. Cosco and M. Borrelli, F. Plastina, and S.  Maniscalco, "Momentum-resolved and correlation spectroscopy using quantum probes", Phys. Rev. A 95, 053620 (2017).

\bibitem{Burgarth} Daniel Burgarth et al, ``Can Decay Be Ascribed to Classical Noise?",Open Syst. Inf. Dyn. 24, 1750001 (2017).


\bibitem{Diosi} Andras Bodor, Lajos Diosi, Zsofia Kallus, and Thomas Konrad, ``Structural features of non-Markovian open quantum systems using quantum chains", Phys. Rev. A 87, 052113 (2013).

\bibitem{Strunz} Silvan Kretschmer, Kimmo Luoma and Walter T. Strunz, ``Collision model for non-Markovian quantum dynamics", Phys. Rev. A 94, 012106 (2016).

\bibitem{Giovannetti} F. Ciccarello, G. M. Palma, V. Giovannetti, ``Collision-model-based approach to non-Markovian quantum dynamics", Phys. Rev. A 87, 040103(R) (2013).

\bibitem{McCloskey} Ruari McCloskey and Mauro Paternostro, ``Non-Markovianity and system-environment correlations in a microscopic collision model", Phys. Rev. A 89, 052120 (2014). 

\bibitem{Nadja} Nadja K. Bernardes, Alvaro Cuevas, Adeline Orieux, C. H. Monken, Paolo Mataloni, Fabio Sciarrino and Marcelo F. Santos, ``Experimental observation of weak non-Markovianity", Scientific Reports 5, Article number: 17520 (2015).

\bibitem{Nadja2} N. K. Bernardes, A. R. R. Carvalho, C. H. Monken, and M. F. Santos, ``Environmental correlations and Markovian to non-Markovian transitions in collisional models", Phys. Rev. A 90, 032111 (2014).

\bibitem{Nadja3} Nadja K. Bernardes, Andre R. R. Carvalho, C. H. Monken, and Marcelo F. Santos, ``Coarse graining a non-Markovian collisional model", Phys. Rev. A 95, 032117 (2017).

\bibitem{Osellame} Jiasen Jin, Vittorio Giovannetti, Rosario Fazio, Fabio Sciarrino, Paolo Mataloni, Andrea Crespi, and Roberto Osellame, ``All-optical non-Markovian stroboscopic quantum simulator", Phys. Rev. A 91, 012122 (2015).

\bibitem{Mesco} M.B. Plenio and S. Virmani. ``Spin chains and channels with memory", Phys. Rev. Lett. 99, 120504 (2007).

\bibitem{Mesco2} M.B. Plenio and S. Virmani, ``Critical phenomena and the capacity of quantum channels with memory", New J. Phys. 10, 043032 (2008).

\bibitem{breuerbook} Heinz-Peter Breuer and Francesco Petruccione, ``The Theory of Open Quantum Systems", Oxford (2007).

\bibitem{Vegathermo} de Vega, In\'es and Ba\~nuls, Mari-Carmen, ``Thermofield-based chain-mapping approach for open quantum systems", Phys. Rev. A 92, 052116 (2015).

\bibitem{Schollwoeck} Ulrich Schollwoeck, The density-matrix renormalization group in the age of matrix product states, Annals of Physics 326, 96 (2011).

\bibitem{Verstraete} F. Verstraete, J.I. Cirac, V. Murg, ``Matrix Product States, Projected Entangled Pair States, and variational renormalization group methods for quantum spin systems", Adv. Phys. 57,143 (2008).


\bibitem{1DBHDMRG} T. D. K�hner and H. Monien, ``Phases of the one-dimensional Bose-Hubbard model", Phys. Rev. B 58, R14741(R) (1998).

\bibitem{Carr} L. D. Carr, M. L. Wall, D. G. Schirmer, R. C. Brown, J. E. Williams, and Charles W. Clark, ``Mesoscopic effects in quantum phases of ultracold quantum gases in optical lattices", Phys. Rev. A 81, 013613 (2010).

\bibitem{WEVs} T. J. Elliott and T. H. Johnson, ``Nondestructive probing of means, variances, and correlations of ultracold-atomic-system densities via qubit impurities", Phys. Rev. A 93, 043612 (2016).

\bibitem{WEVs2} Michael Streif, Andreas Buchleitner, Dieter Jaksch, and Jordi Mur-Petit, ``Measuring correlations of cold-atom systems using multiple quantum probes", Phys. Rev. A 94, 053634 (2016).

\bibitem{Subir} Subir Sachdev and Bernhard Keimer, ``Quantum criticality",  Physics Today 64, 2, 29 (2011).

\bibitem{Coleman} Piers Coleman and Andrew J. Schofield, ``Quantum criticality", Nature 433, 226-229 (2005).

\bibitem{QNoise}  C. W. Gardiner and Peter Zoller, ``Quantum Noise", Springer-Verlag (1991, 2000, 2004).


\bibitem{Nov3} Adrian A. Budini, ``Quantum systems subject to the action of classical stochastic fields", Phys. Rev. A 64, 052110 (2001).


\bibitem{GA} David E. Goldberg, ``Genetic Algorithms in Search, Optimization and Machine Learning", Addison-Wesley, 1989.

\bibitem{Jaksch} Jaksch, D. and Briegel, H.-J. and Cirac, J. I. and Gardiner, C. W. and Zoller, P., "Entanglement of Atoms via Cold Controlled Collisions", Phys. Rev. Lett., 82, 1975 (1999).

\bibitem{bloch2003} Immanuel Bloch, Markus Greiner, Olaf Mandel and Theodor W. Haensch, T. W., "Coherent cold collisions with neutral atoms in optical lattices", Phil. Trans. R. Soc. Lond. A, 361, 1409 (2003).

\bibitem{recati2005} Recati, A. and Fedichev, P. O. and Zwerger, W. and von-Delft, J. and Zoller, P., "Atomic Quantum Dots Coupled to a Reservoir of a Superfluid Bose-Einstein Condensate", Phys. Rev. Lett., 94, 040404, (2005).

\bibitem{devega2008} de Vega, In\'es and Porras, Diego and Ignacio Cirac, J., "Matter-Wave Emission in Optical Lattices: Single Particle and Collective Effects", Phys. Rev. Lett., 101, 260404 (2008).

\bibitem{marino2017} Marino, Jamir and Recati, Alessio and Carusotto, Iacopo, "Casimir Forces and Quantum Friction from Ginzburg Radiation in Atomic Bose-Einstein Condensates", Phys. Rev. Lett. 118, 045301 (2017).

\end{thebibliography}
\end{document}